\documentclass[%
 reprint,
 amsmath,amssymb,
 aps,
]{revtex4-2}

\usepackage{graphicx}
\usepackage{dcolumn}
\usepackage{bm}

\usepackage[normalem]{ulem}
\usepackage{xcolor}

\newcommand{\bref}[1]{(\ref{#1})}
\DeclareRobustCommand{\Erase}{\bgroup\markoverwith{\textcolor{red}{\rule[.5ex]{2pt}{0.4pt}}}\ULon}

\begin{document}

\preprint{APS/123-QED}

\title{(Dis)continuous buckling transition in elastic shell mediated by contact}

\author{Takara Abe}
\affiliation{School of Integrated Design Engineering, Graduate School of Science and Technology, Keio University, 3-14-1 Hiyoshi, Yokohama, Kanagawa, 2238522, Japan.}

\author{Tomohiko G. Sano}%
\email{sano@mech.keio.ac.jp}
\affiliation{Department of Mechanical Engineering, Faculty of Science and Technology, Keio University, 3-14-1 Hiyoshi, Yokohama, Kanagawa, 2238522, Japan.
}%
\affiliation{School of Integrated Design Engineering, Graduate School of Science and Technology, Keio University, 3-14-1 Hiyoshi, Yokohama, Kanagawa, 2238522, Japan.}

\date{\today}

\begin{abstract}
Snap-buckling is a rapid shape transition in slender structures, appearing as a fundamental switching mechanism of natural and man-made systems. Boundary conditions of structures are crucial to predict and control their snap-buckling behavior. However, the general framework that relates boundary conditions, geometry, and performance of structures is still absent to date. 
Here, we study the snap-buckling of hemispherical shells in contact with rigid cylinders of different diameters to uncover the roles of boundary conditions in the dynamic performance of shells. Specifically, we analyze the jumping dynamics of the pneumatically inverted shells placed on the rigid cylinder by combining experiments and analytical theory. We find the characteristic diameter classifying the snap-buckling and jumping dynamics into two: (i)~if the diameter of the cylinder is sufficiently larger than the characteristic diameter, the shell is regarded as in contact with the infinitely large flat plate, (ii)~if not, the cylinder is regarded as a point. The analytical predictions for the jumping performance of the shell supplemented with the characteristic diameter are in excellent agreement with our experimental results. Our study clarifies that contact geometry is crucial in predicting the pathway of snap-buckling, indicating that the dynamic performance of soft robots would be optimized by tuning their surface geometry.

\end{abstract}

\maketitle

\section{Introduction}

Snap-buckling is one of the bifurcation phenomena in a slender structure that transits from one stable state to the other~\cite{Strogatz2014, Audoly2010, Bigoni2015}. The stored elastic energy of rods, plates, or shells is converted into kinetic energy, where rapid and large deformation, together with impulsive force or snapping sounds, follows the change of the control parameters. The snap-buckling phenomena, which are found in nature and engineering applications, result from a complex interplay of elasticity and geometry of structures. For example, the geometry of structures~\cite{Pandey2014} and shells~\cite{Pogorelov1988, Gomez2016, Baumgarten2019}, imperfections~\cite{Qiao2020}, contact friction~\cite{Sano2017}, symmetry of boundary conditions~\cite{Gomez2017, Sano2018, Radisson2023}, twist-bend coupling~\cite{Sano2019}, elasto-magnetic coupling~\cite{Abbasi2023}, viscoelasticity~\cite{Gomez2018PhD, Gomez2019, Liu2021}, constraints~\cite{Armanini2019}and boundary  conditions~\cite{Huang2024} influence their buckling conditions and post-buckling behaviors. The tunable and predictive switching mechanism of snap-buckling is found in nature, engineering designs, and children's toys, \textit{e.g.,} nastic motion in plants~\cite{Dumais2011}, insects\cite{Wang2023}, sensors~\cite{Medina2014}, soft actuators~\cite{Rus2015}, and popping toys~\cite{Lapp2008, Ucke2009}.

\begin{figure}[!b]
\includegraphics[width =0.48\textwidth]{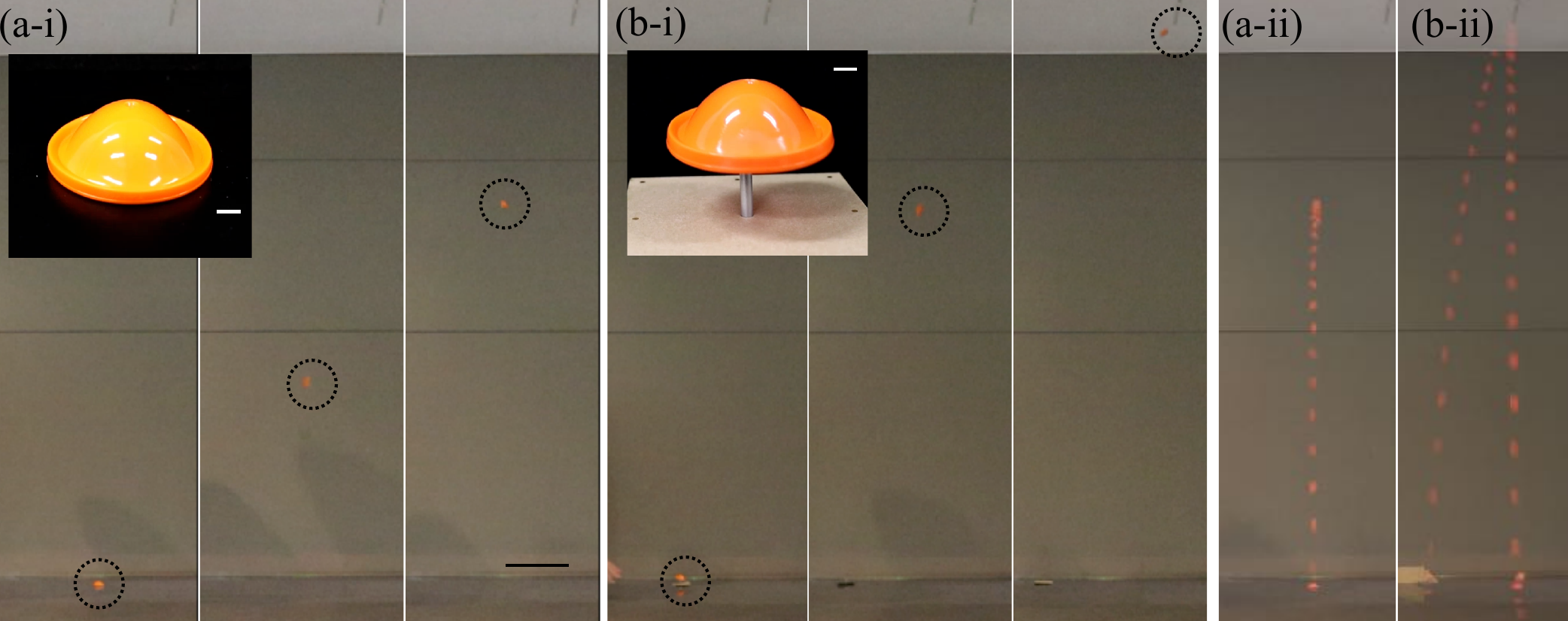}
\caption{The demonstration of the jumping experiment of popper toys on (a)~the flat substrate (floor) and (b)~rigid protrusion (6~mm screw). (b-i)~The popping on the protrusion reaches the ceiling, not (a-i)~on the flat substrate (scale bar 30~cm). The insets represent the corresponding initial (buckled) configurations (scale bar 10~mm). The multiple exposures of the popping phenomena (available as Supplementary Movie S1) on (a-ii) flat and (b-ii) protrusion. Note that all the experimental data analyzed in this paper are obtained in the jumping experimental system in Fig.~\ref{Fig2}.}
\label{Fig1}
\end{figure}

The jumping or eye popper toy is a children's toy made of an elastomeric hemispherical shell. The inverted shell flips and jumps, followed by snap-buckling instability~\cite{Lapp2008, Ucke2009, Pandey2014}. The inverted shell is usually placed on a flat substrate, and it pops out with a snapping sound. When we place it on a rigid protrusion, the shell reaches a much higher height than the flat substrate (See Fig.~\ref{Fig1}). Although this remarkable difference in the jumping performance of the popper highlights the fact that the boundary condition (contact between shell and substrate) is crucial in forecasting snap-buckling instability, the general framework for the jumping behavior of the popper has not been investigated thus far~\cite{Lapp2008, Ucke2009}. In the previous work, we propose the analytical framework against the jumping performance of the pneumatically-controlled popper on a flat substrate, which is consistent with experiments and simulations~\cite{Abe2025}. Our experimental system allows us to study the roles of shell geometry, material stiffness, and net weight systematically (detailed briefly in Sec.~\ref{sec:def}). Nevertheless, the role of the substrate geometry, i.e., \textit{why does the popper jump higher on the protrusion}, is not clarified to date.

The fundamental aspect of snap-through dynamics has been studied within the context of the mechanics of materials recently~\cite {Gomez2018PhD, Gomez2019}. For example, Gomez, \textit{et al.,} formulates the viscoelastic snap-through dynamics of arches, employing the analysis of the von Mises truss~\cite{Gomez2019}. They predict the timescale for the snap-through dynamics, highlighting the effect of the viscoelasticity of materials. Although several features of snap-through dynamics have been applied to engineering problems, \textit{e.g,}, Micro Electro Mechanical Systems (MEMS)~\cite{Zhang2007}, shock protections~\cite{Noh2025}, soft robots~\cite{Gorissen2020}, the prediction of their performances are challenging due to inherent coupling between material properties and nonlinear geometry.

Here, we systematically study the roles of substrate geometry in the jumping performance of the elastomeric shell. We experimentally analyze the jumping process of the shell on the rigid cylinder of the small diameter $d$. We propose the predictive framework for the jumping performance on the narrow substrate by applying the analytical formula for the point-indentation test of the shell. The analytical formula is derived and compared with the experimental results, and it is in excellent agreement with experimental data without any fitting parameters. We find that when the diameter is sufficiently larger than the threshold, $d^*$, the jumping performance converges to that on the flat substrate reported in the previous study~\cite{Abe2025}, whereas, in the other limit, $d\ll d^*$, the jumping height becomes large and is predicted by the analytical formula. In other words, we fully elucidate the roles of substrate geometry in the jumping phenomena of the popper. 
Our finding would be applicable in predicting the performance of the jumping robot in complex or irregular environments.

The structure of this paper is as follows. In Sec.~\ref{sec:def}, we define the problem and introduce experimental setups. The fabrication of the elastomeric shells and the experimental procedures are detailed. In Sec.~\ref{sec:theory}, we propose analytical formulas for the jumping performance of the popper on the flat or point-like substrates. We also validate the analytical force-displacement curves derived in the literature~\cite{Audoly2010, Baumgarten2019} against our experimental samples. In Sec.~\ref{sec:comparison}, we compare our analytical formula and experimental results. We analyze the maximum jumping height and the jumping condition of the popper on the point-like substrate. In addition, we show that the jumping height increases as $d^*/d$ with the logistic curve-like, indicating that there exists a critical diameter, $d^*$, that the popper can jump much higher than on the flat substrate. Lastly, in Sec.~\ref{sec:dis}, we conclude the paper.

\section{Sample fabrication and Setup of jumping experimental system}\label{sec:def}

\begin{figure}[!h]
\includegraphics[width =0.5\textwidth]{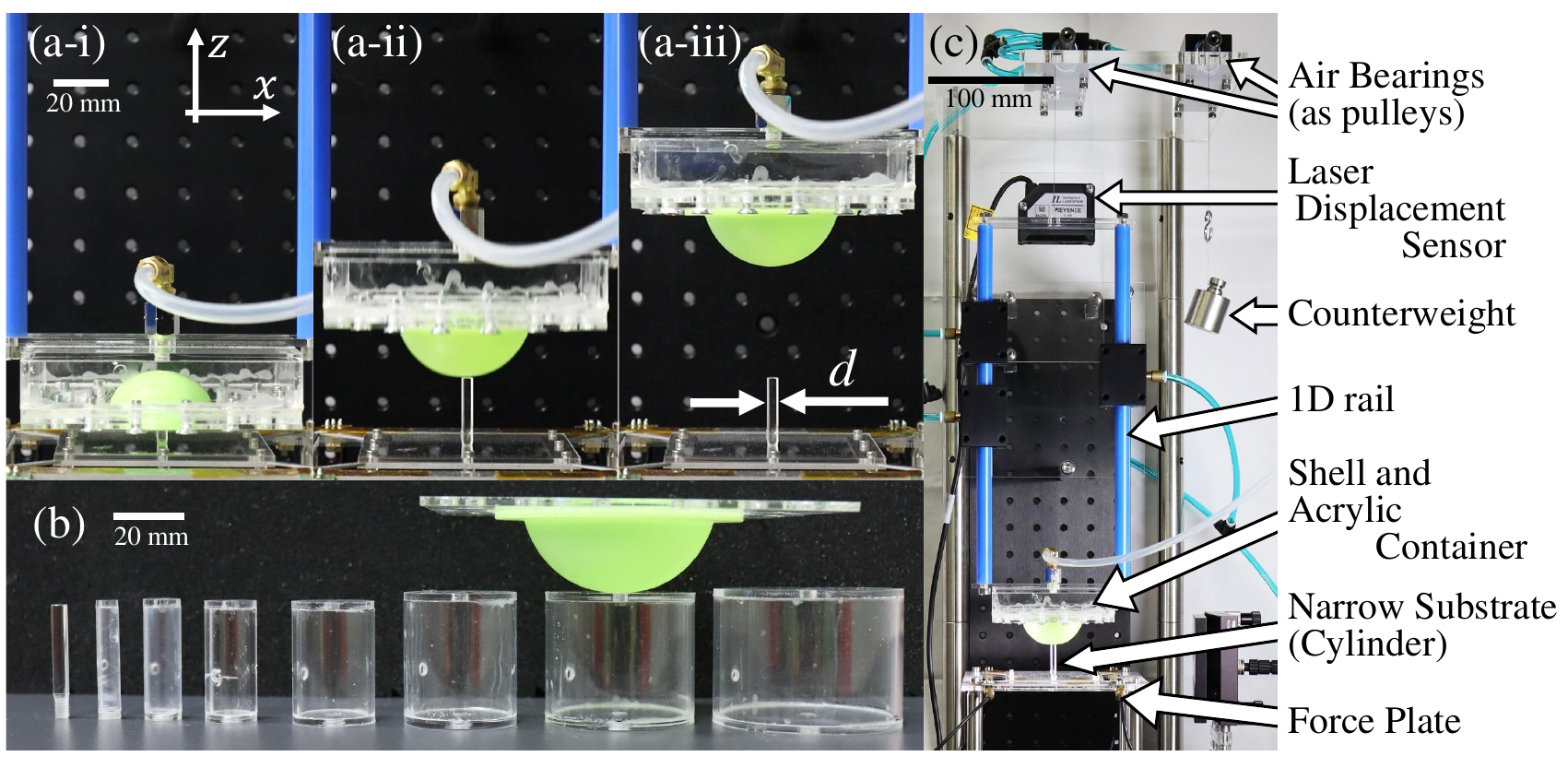}
\caption{The pneumatically-controlled elastic shell jumps on a narrow substrate.
(a-i)-(a-iii)~The front view of the typical jumping process of our elastic spherical shell on a narrow substrate ($(R, h)= (25, 2.0)$~mm, $d = 4$~mm).
(b)~Photographs of narrow substrates (Acrylic cylinders) with different diameters $d \in [4, 50]$~mm.
(c)~Photograph of the full experimental apparatus. 
The set of the container and shell (popper) rises vertically along one-dimensional rails. The popper is connected to the counterweight via air bearings. In each jumping test, the vertical position, pressure of the container, and normal reaction force of the substrate are measured simultaneously by combining the laser displacement sensor, the pressure sensor, and the custom-made force sensor plate connected with the cylinder. Jumping experiments are performed on different substrates by replacing the cylinder on the force plate.}
\label{Fig2}
\end{figure}

In this section, we describe the jumping experimental system initially proposed in Ref.~\cite{Abe2025}. We fabricate elastomeric shells made of silicone elastomer (Elite Double (ED) 22 and ED32, Zhermack, Italy) with homogeneous thickness, $h$, the radius of curvature $R$, and the polar angle, $\varphi$. The viscoelasticity of this elastomer is sufficiently small, such that the mechanical performances of our samples could be modeled as incompressible hyperelastic materials of Young's modulus, $E$ (Poisson's ratio $\nu = 0.5$)~\cite{Grandgeorge2021}. The base and catalyst are mixed in the centrifugal mixer (AR-100, Thinky Corporation, Japan) and degassed in the vacuum chamber to reduce possible defects by removing air bubbles. The mixed liquid elastomer is injected into the 3D-printed mold and is cured at room temperature. After curing ($\sim30$~min.), the shells are demolded and glued to the acrylic container (See Fig.~\ref{Fig2}(a,b)). We systematically fabricate shells of different sizes and aspect ratios, fixing the polar angle, $\varphi$, as $\varphi = 80^{\circ}$ throughout. 

The set of shell and container, which we call \textit{popper} here, is fixed with two rigid shafts acting as one-dimensional (1D)-rails. The rigid shafts can slide along air bearings (13mm ID Air Bushing, Newway, USA) smoothly under ideally low friction. The rigid shafts are connected with the counterweight via a Nitinol wire and air bearings (acting as pulleys) to precisely control the net weight of the popper, $mg$ (Fig.~\ref{Fig2}(c)). The vertical location of the popper, $z$, the pressure difference inside the container from the atmospheric pressure, $\delta p$, and the reaction force, $F$, are measured by the laser displacement sensor (IL-300, KEYENCE, Japan, 3~kHz sampling rate), pressure sensor (AP-C30, KEYENCE, Japan, 400~Hz sampling rate), and the custom-made force plate attached to the strain gauge, respectively. The time series data for $z, \delta p$, and $F$ are time-synchronized to correlate their dynamics. The rigid cylinder of diameter, $d$, and length $30$~mm fixed to the force plate is replaceable (Fig.~\ref{Fig2}(b)). We call the substrate with the cylinder the narrow substrate.

The experimental protocol is identical to our previous work~\cite{Abe2025} and is detailed briefly as follows. The container is depressurized as $\delta p<0$ by a syringe. When the pressure difference exceeds the critical value, the shell is inverted (shell-buckling). The inverted shell and container under the pressure difference, $\delta p_j$, is lowered along the rail and is grounded with the substrate, where the top of the cylinder and the shell apex are in contact. The central axes of the cylinder and shell are set to be aligned coaxially. After grounding the popper, we open the valve connected to the container at time $t = 0$ so that the internal pressure reaches nearly atmospheric pressure. The shell recovers its original shape, and the popper is launched from the initial height at $z = 0$.

\begin{figure}[!b]
\includegraphics[width =0.5\textwidth]{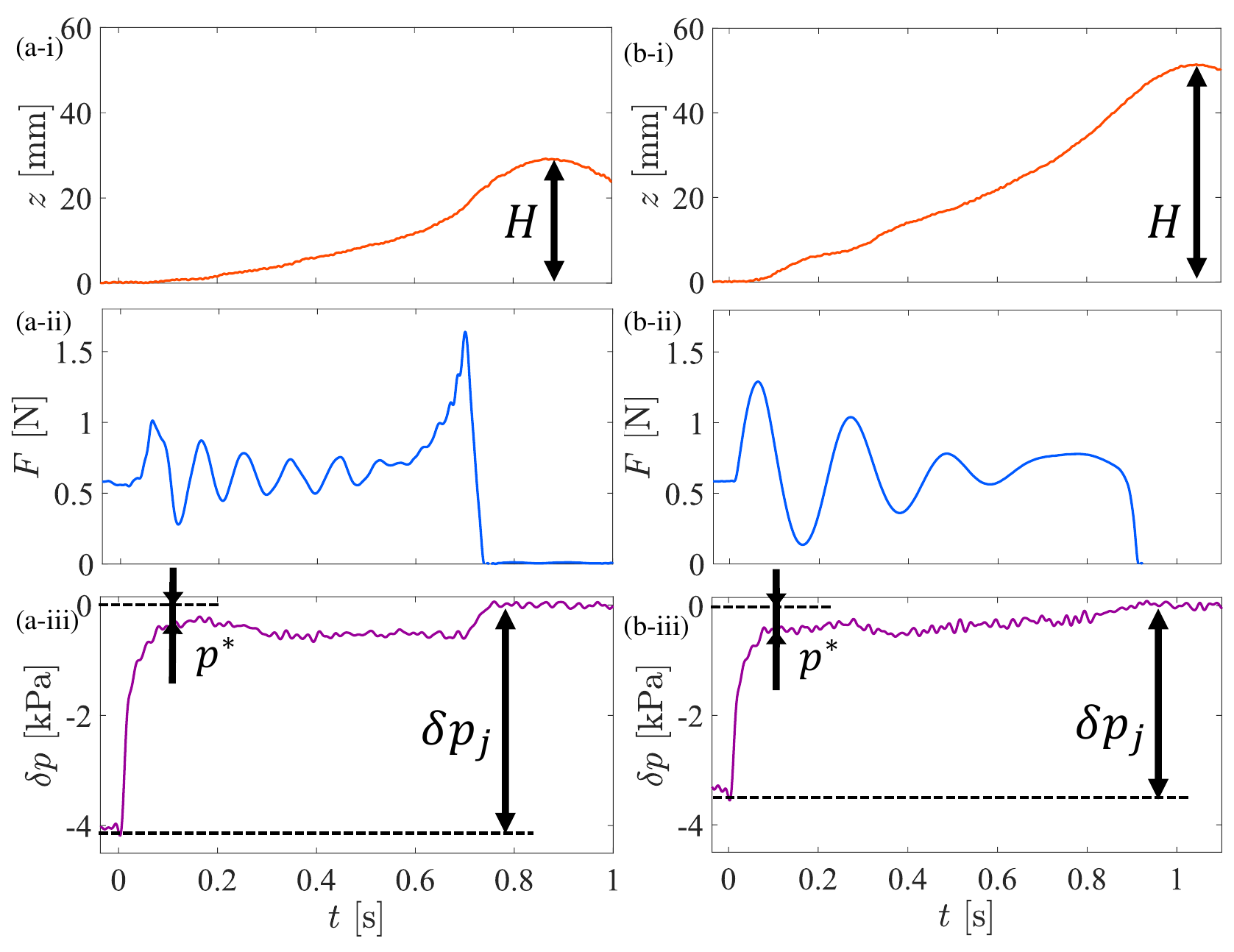}
\caption{
Comparison of the jumping phenomena on the (a)~narrow and (b)~flat substrate ($(R, h)= (30, 2.0)$~mm, $mg = 0.6$~N). 
Typical time series data for the jumping process on the flat substrate ((a-i)-(a-iii)), and on the narrow substrate ((b-i)-(b-iii)): (i) the vertical position of the container, $z$, (ii) the reaction force of the substrate $F$, and (iii) the pressure difference between inner and atmospheric pressures, $\delta p$. 
}
\label{Fig3}
\end{figure}

We show the typical time series data for the popper height, $z$, internal pressure, $\delta p$, and the reaction force, $F$ in Fig.~\ref{Fig3}. We compare the results on the flat substrate and those on the cylinder of diameter $d = 6~$mm for $(R, h) = (30, 2.0)$~mm and $mg = 0.6$~N in Fig.~\ref{Fig3}(a) and (b), respectively. When the valve is opened at $t = 0$, the air flows into the container instantaneously, and the shell starts to recover its original shape. The longitudinal wave of the wire connecting the popper and the counterweight results in the damped oscillation right after the valve opens, while pressure saturates to a nearly constant value immediately. The popper is then launched and reaches the maximum height, $H$. The remarkable difference between flat and narrow substrates is the force right before the jump (Fig.~\ref{Fig3}(a, b-ii)). The characteristic impulsive force exists when the popper jumps on the flat substrate, while it does not when the popper launches from the narrow substrate. It should be noted that even if the impulsive force does not exist for the narrow substrate, the popper can jump higher than on the flat substrate (Fig.~\ref{Fig3}(a, b-i)) consistent with our demonstration (Fig.~\ref{Fig1}). Below, we aim to formulate analytical predictions of our jumping performance both for flat and narrow substrates, employing the elasticity and geometry of flat/point indentation tests.


\section{Analytical framework for jumping performance of shells}\label{sec:theory}

We formulate the analytical framework for the jumping experiment, based on the elasticity and geometry of elastic shells~\cite{Audoly2010}. We first summarize the formulas for the jumping performance on the flat substrate validated experimentally and numerically in Ref.~\cite{Abe2025} (Sec.~\ref{sec:flat_theory}). In Sec.~\ref{sec:narrow_theory}, we propose the corresponding formulas for the sufficiently narrow substrate whose diameter is regarded as a point $d/R\to0$. 

\subsection{Jumping on the flat substrate}\label{sec:flat_theory}

When the buckled shell is placed on the flat substrate, it jumps as the shell recovers its original shape, with a snapping sound. In Ref.~\cite{Abe2025}, it is shown that the contact geometry transition, where the ring-like contact shape (via the rim of the buckled shell) transits to the areal disk-like contact. First, we analyze the indentation test of the shell with the flat substrate within the scaling level and validate the predictions experimentally in Sec.~\ref{sec:fd_flat}. The discussion for the contact geometry follows in Sec.~\ref{sec:contact_trans}. Lastly, we formulate the formulas for the maximum height on the flat substrate in Sec.~\ref{sec:height_flat}.

\subsubsection{Indentation test with the flat indentor}\label{sec:fd_flat}

We press the shell by the flat indenter from the apex (in the absence of pressure difference between the outside and inside the shell). 
When the plate indents the shell by the apex displacement, $e$, slightly, the shell contacts the plate via disk (areal) contact of the radius, $r$. The apex displacement, $e$, and the contact radius are related as 
\begin{eqnarray}
r\sim\sqrt{eR}\label{eq:reR}
\end{eqnarray}
by geometry. Relative to the natural length of the contact region of length $R\sin^{-1}(r/R)$, the radial arclength is compressed by $R\sin^{-1}(r/R)-r\simeq r^3/R^2$, from which we estimate the scaling of the strain as $\epsilon\sim (r/R)^2$. 

As we increase the displacement, $e$ (and $r$), both elastic bending and stretch energy increase with the strain, $\epsilon\sim(r/R)^2$, and curvature change (from $1/R$ to 0) over the contact region. The in-plane stretching (compression) energy of the contact area, $\mathcal{E}_{\rm st}$, changes as
\begin{eqnarray}
    \mathcal{E}_{\rm st} \sim \int Eh\epsilon^2 dA \sim \frac{Ehr^6}{R^4},
    \label{eq:st}
\end{eqnarray}
while the bending energy, $\mathcal{E}_{\rm b}$, is
\begin{eqnarray}
    \mathcal{E}_{\rm b} \sim Eh^3 \int \frac{1}{R^2}dA \sim \frac{Eh^3}{R^2}r^2.
    \label{eq:b}
\end{eqnarray}
The energetic cost for stretching is small for slight indentation as $\mathcal{E}_{\rm st}\ll\mathcal{E}_{\rm b}$, where the flat disk contact is realized. As the apex height of the shell is decreased and the contact radius, $r$ increases, the stretching and bending energies (Eqs.~\bref{eq:st},(\ref{eq:b})) become nearly the same order at the characteristic indentation depth, $e_c$, and contact radius $r_c$. Above the characteristic indentation, $e\gg e_c$, the flat contact is less favorable as $\mathcal{E}_{\rm st}\gg\mathcal{E}_{\rm b}$ and the apex flips and snaps, where the bending is localized around the buckled rim. The scaling relations for characteristic indentation depth, $e_c$, and contact radius $r_c$ for the snap transition are obtained by balancing stretching and bending energies, Eqs.~(\ref{eq:st}), (\ref{eq:b}) as
\begin{eqnarray}
    e_c = h,~~r_c =\sqrt{hR}\label{eq:ecrc_s}.
\end{eqnarray}

\begin{figure}[!h]
\includegraphics[width =0.5\textwidth]{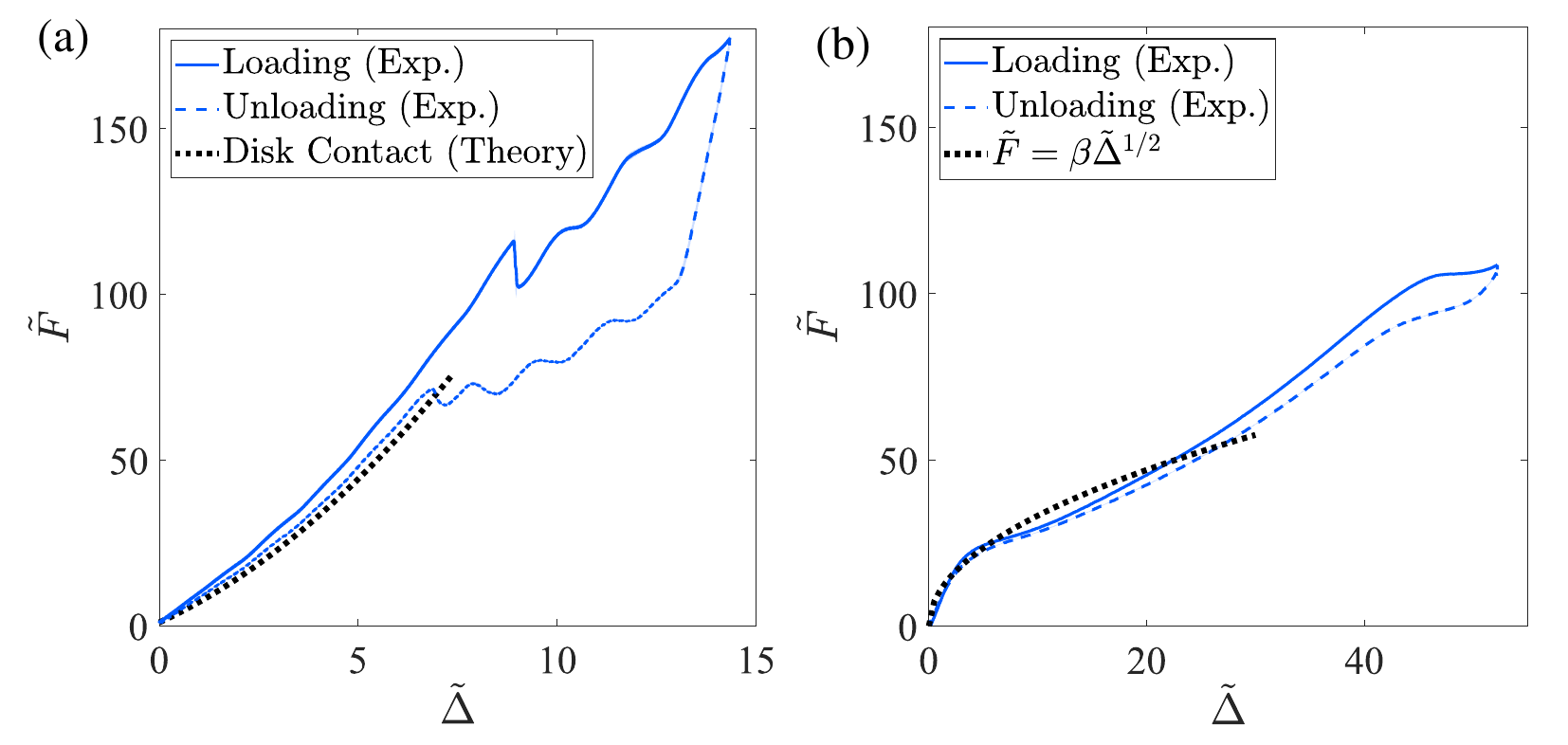}
\caption{
The rescaled force-displacement curves of the shell  ($(R, h)= (25, 2.0)$~mm) for the indentation test using the (a)~flat plate and (b)~narrow cylinder ($d=6$~mm). The blue solid and dashed lines correspond to the experimental loading and unloading processes, respectively.
The black dashed line in (a) represents the analytical theory assuming the disk (areal) contact~\cite{Audoly2010}, while that in (b) is the analytical prediction for the point indentation test~\cite{Baumgarten2019}. The theory has no adjustable parameters.
}
\label{Fig4}
\end{figure}

The indenting force, $F(e)$, increases with the apex displacement, $e$. The scaling for the force in the disk contact, $F_{\rm disk}$, is given by $F\sim\partial\mathcal{E}_{\rm st}/\partial e + \partial\mathcal{E}_{\rm b}/\partial e$, i.e.,
\begin{eqnarray}
    F_{\rm disk} \sim  \frac{Eh^3}{R}\left\{1 + \left(\frac{e}{h}\right)^2\right\}.
\end{eqnarray}
The rigorous analytical prediction for $F = F(e)$ with the disk contact, $F_{\rm disk}$, is numerically obtained as
\begin{eqnarray}
    F_{\rm disk} = \frac{Eh^3}{12(1-\nu^2)R}\hat{N}(\hat{e}),\label{eq:FD_disk}
\end{eqnarray}
with the convex function of the rescaled displacement, $\hat{e}\equiv \sqrt{12(1-\nu^2)}e/h$, $\hat{N} = \hat{N}(\hat{e})$. 
Equation (\ref{eq:FD_disk}), whose functional form can be obtained numerically as detailed in Ref.~\cite{Audoly2010}, correctly predicts the experimental force-displacement curve prior to the snap-buckling threshold by identifying the apex displacement, $e$, as the displacement of the indenter, $\Delta$, as shown in Fig.~\ref{Fig4}(a). The rescaled displacement, $\tilde{\Delta}$, and force, $\tilde{F}$, are introduced as $\tilde{\Delta}\equiv\sqrt{12(1-\nu^2)}\Delta/h$ and $\tilde{F}\equiv F/\{Eh^3/(12(1-\nu^2)R)\}$.  
The characteristic force necessary to snap the shell, $F_c$, is estimated by balancing the work done by the external force, $F_ce_c$, and the stored elastic energy at the transition, $\sim Eh^4/R$ (obtained by substituting Eq.(\ref{eq:ecrc_s}) into Eq.(\ref{eq:b})), i.e., we get
\begin{eqnarray}
    F_c = \frac{Eh^3}{R}.
\end{eqnarray}

Above the snap-buckling threshold ($F\gg F_c$), the force-displacement curve transits to that for the ring contact, $F_{\rm ring}$, given by
\begin{eqnarray}
    F_{\rm ring} = \frac{Eh^3n_{(2)}}{\{12(1-\nu^2)\}^{3/4}R} \sqrt{\frac{e}{h}},\label{eq:FD_ring}
\end{eqnarray}
with the dimensionless prefactor derived as $n_{(2)}\simeq20.83$~\cite{Audoly2010}. It should be noted that $F_{\rm ring}$ is described as a function of the apex displacement, $e$, not the displacement of the indenter, $\Delta$ in Eq.~(\ref{eq:FD_ring}). In other words, one may need to relate $e$ and $\Delta$ to compare with the experimental force-displacement curve for the ring contact. 

We observe the hysteresis in the force-displacement curve for the plate indentation (Fig.~\ref{Fig4}(a)) as discussed in Ref.~\cite{Abe2025}. The existence of the hysteresis is crucial in predicting the jumping dynamics, which will be detailed in the next section.

\subsubsection{Contact geometry transition}\label{sec:contact_trans}

The contact geometry of the shell-indenter shifts between disk and ring states during loading and unloading protocols, respectively. The exact transition points are obtained through a stability analysis of the shell configurations associated with either disk or ring contact. 
According to the previous study~\cite{Audoly2010}, the disk contact configuration in the loading process becomes linearly unstable above the critical force satisfying $\hat{N} = N_+\simeq 76.94$ in Eq.~(\ref{eq:FD_disk}). The ring contact configuration becomes unstable upon decreasing the force. The minimum possible values of the critical force for the ring-to-disk transition are computed by finding the root of $F_{\rm ring} = F_{\rm disk}$ as $\hat{N}= N_-\simeq46.8$~\cite{Abe2025}. The contact geometry between the shell and the flat plate is categorized into three: (i)~The contact shape is disk-like for $F<F_-\equiv Eh^3N_-/\{12(1-\nu^2)\}$, (ii)~The contact shape is either disk or ring-like for $F_-<F<F_+\equiv Eh^3N_+/\{12(1-\nu^2)\}$, depending on the loading/unloading protocols, (iii)~The contact shape is ring-like for $F>F_+$.

Using the characteristic force scale, $F_c$, the critical force for the contact geometry transition, $F^*$, is described as
\begin{eqnarray}
    F^* = \alpha_F F_c,\label{eq:Fc}
\end{eqnarray}
where the dimensionless prefactor, $\alpha_F$, satisfies $F_-/F_c\leq\alpha_F\leq F_+/F_c$ ($F_-/F_c\simeq5.3$ and $F_+/F_c\simeq8.5$). 
Other critical physical quantities for the transition (\textit{e.g.,} critical indentation depth, $e^*$ and contact radius, $r^*$) are computed numerically based on the above considerations with the aid of dimensionless prefactors bounded by the coexistence of the ring and disk contact geometry for a given loading force in $F_-<F<F_+$;
\begin{eqnarray}
    e^* = \alpha_e e_c,~~r^* = \alpha_r r_c\label{eq:ecrc},
\end{eqnarray}
with $1.8\leq\alpha_e\leq4.5$ and $1.2\leq\alpha_r\leq1.6$. Equations (\ref{eq:Fc}) and (\ref{eq:ecrc}) are validated experimentally and numerically for several sets of $(h, R)$ in Ref.~\cite{Abe2025}.

The jumping process of the elastic shell on the flat substrate utilizes the contact geometry transition, where the contact shape transits from ring to disk (following the lower branch of Fig.~\ref{Fig4}(a)). By dividing the jumping process into pre- and post-transition regimes, the maximum jumping height on the flat substrate is derived in Ref.~\cite{Abe2025}. We briefly summarize the prediction in the next subsection. 

\subsubsection{The maximum jumping height on the flat substrate}\label{sec:height_flat}

On the basis of the force-displacement curve for the plate indentation of the shell, we now construct the analytical formula for the maximum jumping height on the flat substrate. The idea to derive the formula is that the jumping process can be regarded as the unloading process for the indentation test (Fig.~\ref{Fig4}(a)).

We divide the jumping process into the pre- and post-contact geometry transition regimes, i.e.,  Before the contact geometry transition, the shell is in contact with the substrate via the rim of the buckled shell. The popper is \textit{lifted} slowly in this regime, allowing us to estimate the displacement of the popper geometrically, $H_{\rm lift}$, as
\begin{eqnarray}
   H_{\rm lift} \simeq R(1-\cos\varphi) - e^*.
   \label{eq:Hlift}
\end{eqnarray}
Here, we have assumed that the popper is sufficiently depressurized such that the shell is mirror-buckled, where the difference in the height of the buckled rim and that of the apex is approximately given as $R(1-\cos\varphi)$. 
Equation~(\ref{eq:Hlift}) is consistent with experimental and numerical results for the sufficiently inverted shells. 

When the apex displacement reaches critical, $e^*$, the contact geometry transits to the disk whose static configurations are characterized by $r^*, e^*$ and $F^*$. The elastic energy stored in the critical disk contact configuration is estimated as $\mathcal{E}^* = F^*e^*$, leading to the analytical prediction for the maximum popper height in the post-transition regime, $H_{\rm s}$, as, $\mathcal{E}^* = mgH_{\rm s}$, i.e.,
\begin{eqnarray}
    H_{\rm s} = \alpha_F\alpha_e \frac{Eh^4}{mgR}. 
    \label{eq:Hs}
\end{eqnarray}
Combining Eqs.~(\ref{eq:Hlift}) and (\ref{eq:Hs}), we can construct the analytical formula for the maximum jumping height on the flat substrate, $H_{\rm flat}$, as
\begin{eqnarray}
    H_{\rm flat} &=& H_{\rm lift} + H_{\rm s}\nonumber\\
    &=& R(1-\cos\varphi) - \alpha_eh + \alpha_F\alpha_e \frac{Eh^4}{mgR}\label{eq:Hflat}.
\end{eqnarray}
In summary, the jumping performance of shells on the flat substrate is entirely predictable with the aid of the elasticity and geometry of shells. In the next section, we construct the analytical prediction when the shell jumps on a sufficiently narrow substrate to clarify the roles of substrate geometry in the jumping process.  

\subsection{Jumping on the narrow substrate $d/R\to0$}\label{sec:narrow_theory}

The shell jumps smoothly on the sufficiently narrow substrate without snapping sounds or impulsive forces. The difference from the flat substrate originates from the fact that the contact geometry transition is now prohibited due to the rigid cylinder attached to the substrate. The buckled radius decreases continuously to zero during the jumping process, where the apex stays in contact with the substrate, following Eq.~(\ref{eq:reR}), and then the popper leaves the substrate. We construct the analytical formula for the maximum jumping height based on the scenario without contact transition. In Sec.~\ref{sec:fd_point}, we review the previous works for the indentation test for the spherical shells based on shallow-shell theory~\cite{Audoly2010, Baumgarten2019}. 

\subsubsection{Point indentation test}\label{sec:fd_point}

Suppose that the shell is point-indented at the apex with the displacement, $e$. For simplicity, the pressure difference between inside and outside the shell is set to zero, the effects of which will be considered later. The shell buckles with the rim-radius $r$ satisfying the relation same as Eq.~\eqref{eq:reR}, $r\sim\sqrt{eR}$, and with the ridge of finite width $\delta$. The ridge width decreases with the shell thickness, $h$,~\cite{Pogorelov1988, Audoly2010, Gomez2016}. The ridge has two principal curvatures: the radius of the meridian, $R_1\sim\delta/\theta$, and that of the latitude, $R_2\sim R (\gg R_1)$, with the kink angle $\theta\sim e/r\sim\sqrt{e/R}$ (angles formed by longitudinal tangents of deformed and natural configurations). The Gauss curvature defined as $\kappa_G \equiv 1/(R_1R_2)$ is scaled as $\kappa_G\sim \theta/(R\delta)$, giving the scaling for the stretch as $\epsilon\sim\delta^2\kappa_G\sim\theta\delta/R$ (from the compatibility relation~\cite{Audoly2010}). 

The ridge width, $\delta$, is determined by the balance of the stretching and bending energy as we obtain the characteristic length scales for the flat substrate case (Sec.~\ref{sec:fd_flat})~\cite{Audoly2010}. The stretching, $\varepsilon_{\rm st}$ and bending energy per unit area, $\varepsilon_{\rm b}$, are respectively given by
\begin{eqnarray}
    \varepsilon_{\rm st} &\sim& Eh\epsilon^2\sim Eh\frac{\theta^2\delta^2}{R^2},\\
    \varepsilon_{\rm b} &\sim& \frac{Eh^3}{R_1 ^2}\sim Eh^3 \frac{\theta^2}{\delta^2}.
\end{eqnarray}
By balancing above two energies as $\varepsilon_{\rm st}\sim\varepsilon_{\rm b}$, we obtain the scaling relation for the ridge width, $\delta$, as
\begin{eqnarray}
    \delta\sim\sqrt{hR}.\label{eq:ridge}
\end{eqnarray}
Substituting all the scaling predictions into the elastic energy and assuming the area of the ridge region is estimated as $\sim2\pi r\delta\sim2\pi R\sqrt{eh}$, we obtain the elastic energy for the point-indentation~\cite{Audoly2010}:
\begin{eqnarray}
    \mathcal{E}_{\rm point} \sim E\frac{h^{5/2}e^{3/2}}{R}.\label{eq:Ep}
\end{eqnarray}
The scaling relation for the indentation force, $F_{\rm point}$, is now obtained by differentiating Eq.~\eqref{eq:Ep} with respect to $e$, $F_{\rm point}\sim\partial \mathcal{E}_{\rm point}/\partial e$, i.e.,
\begin{eqnarray}
    F_{\rm point} \sim \frac{Eh^{5/2}}{R}e^{1/2}.
\end{eqnarray}
The rigorous expression for the elastic energy and the indenting force are respectively derived in the previous literature~\cite{Baumgarten2019} as
\begin{eqnarray}
    \mathcal{E}_{\rm point} &=&\frac{2\beta}{3} \frac{Eh^{5/2}}{\{12(1-\nu^2)\}^{3/4}R} e^{3/2}\label{eq:Ep}\\
    F_{\rm point} &=& \beta \frac{Eh^{5/2}}{\{12(1-\nu^2)\}^{3/4}R} e^{1/2}\label{eq:Fp},
\end{eqnarray}
with the dimensionless prefactor $\beta = 10.48$. 
We compare the previous analytical result Eq.~(\ref{eq:Fp}) against the experimental data for the indentation test by identifying $e$ by the displacement of the load cell, $\Delta$ in Fig.~\ref{Fig4}(b). In contrast to the plate indentation, there is almost no hysteresis between the loading and unloading processes. Equation~(\ref{eq:Fp}) is in excellent agreement with the experimental data in the range of $\tilde{\Delta}\lesssim30$. Although it would be necessary to extend Eqs.~(\ref{eq:Ep}) and (\ref{eq:Fp}) to derive the more predictive formula of $F_{\rm p}$ for $\tilde{\Delta}\gtrsim30$~\cite{Audoly2010}, Equations~(\ref{eq:Ep}) and (\ref{eq:Fp}) provide sufficiently accurate analytical formula in our jumping experiments as detailed later. The systematic derivations for Eqs.~(\ref{eq:Ep}) and (\ref{eq:Fp}) is detailed in Ref.~\cite{Baumgarten2019}.

In Ref.~\cite{Baumgarten2019}, the energy landscape for the indentation is further derived when the shell is pressurized from inside by pressure, $p$. The contribution from the pressure difference is linearly added to Eqs.~(\ref{eq:Ep}) and (\ref{eq:Fp}) by taking into account the volume change, $\sim \pi e^2 R/2$. The necessary work to indent the shell by the displacement, $e$, under the compressive pressure, $p$ (the difference of outer pressure from inner pressure), $\mathcal{E}_{\rm point}(e,p)$, is represented as
\begin{eqnarray}
    \hspace{-0.4cm}
    \mathcal{E}_{\rm point}(e,p) = \frac{2\beta}{3} \frac{Eh^{5/2}}{\{12(1-\nu^2)\}^{3/4}R} e^{3/2} - \frac{\pi}{2}e^2Rp.\label{eq:Epp} 
\end{eqnarray}
We use the above analytical formula Eq.~\eqref{eq:Epp} to predict the jumping behavior of the popper. 

\subsubsection{Formulas for the maximum jumping height on the narrow substrate}\label{sec:height_narrow}

On the basis of the point-indentation response of the shell, we now formulate the jumping performance of the popper. Specifically, we aim to calculate the maximum jumping height of the popper launched from the point-like (narrow) substrate, $H_{\rm point}$. In contrast to the flat substrate, the shell apex is always in contact with the narrow substrate before the launch, i.e., their contact geometry remains unchanged, and the shell does not pass through the contact geometry transition. Hence, we expect that the maximum jumping height is derived by assuming that the necessary work to indent the shell (stored total energy of the system), $\mathcal{E}_{\rm point}$, is fully converted into the gravitational potential energy, $mgH_{\rm point}$. 

In addition to the shell elastic energy, we would need to consider the work done by the internal air because the shell deforms ($e$ decreases) against the pressure difference. We relate the compressive pressure, $p^*$ (See Fig.~\ref{Fig3}(b)), by the force balance applying the formulas in Sec.~\ref{sec:fd_point}. Given that the air is discharged slowly (as seen in the time series data of Fig.~\ref{Fig3}(b)), one can assume that the indentation force, distributed load from the pressure difference, and the net weight are balanced, i.e., we expect the balance equation in the buckled region as
\begin{eqnarray}
    F_{\rm point}(e) - \pi r^2 p^* = mg\label{eq:pst},
\end{eqnarray}
which allows us to describe the pressure, $p^*$, as a function of $e$ as $p^* = p^*(e)$. By substituting $p^*(e)$, into the energy conservation law as $\mathcal{E}_{\rm point}(e,p^*(e)) = mgH$, we obtain the analytical prediction for the maximum jumping height as
\begin{eqnarray}
    H_{\rm point} = \frac{\beta}{6\{12(1-\nu^2)\}^{3/4}}\frac{Eh^{5/2}e^{3/2}}{mgR} + \frac{e}{2}. \label{eq:Hp}
\end{eqnarray}
Equation (\ref{eq:Hp}) provides the maximum jumping height of the popper indented by $e$. It should be noted that $H_{\rm point}$ is not proportional to the inverse of the weight, $1/mg$. Their counterintuitive relation is originated from the complex interplay of the internal air and shell elasticity formulated in Eq.~\eqref{eq:pst}. In the next section, we validate the above predictions through our experimental setup.

\section{Comparison with experiment}\label{sec:comparison}

We have formulated the jumping performance of the shell placed on the sufficiently narrow substrate ($d\to0$ limit). We now validate our predictive framework in two folds. In Sec.~\ref{sec:Hn}, we measure the maximum jumping height and compare its result against our analytical prediction, Eq.\eqref{eq:Hp}. In Sec,~\ref{sec:JumpNot}, by applying the formula, we predict the jumping condition of the popper. Lastly, in Sec.~\ref{sec:dia}, we discuss when the cylinder diameter is regarded as a point or a flat substrate based on the theoretical formulations detailed in Sec.~\ref{sec:theory}.

\begin{figure}[!h]
\includegraphics[width =0.5\textwidth]{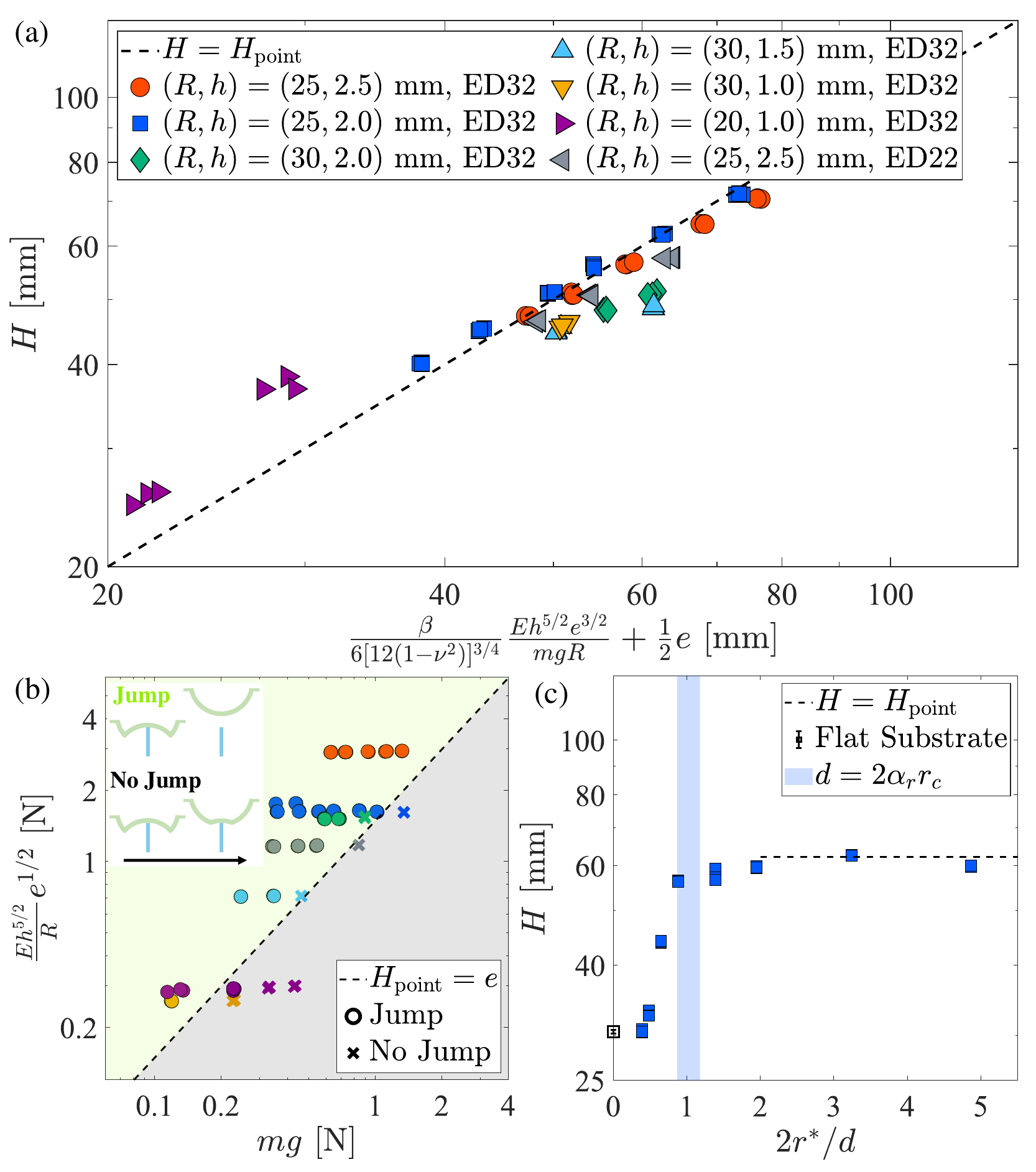}
\caption{Jumping performance of elastic shells on the narrow substrates.
(a)~The jumping height of the shell $H$ on the narrow substrate ($d = 6$~mm) plotted against the analytical formula (dashed line) Eq.(\ref{eq:Hp}). 
(b)~{The phase diagram for the jump on the narrow substrate ($d = 6$~mm). The dashed line represents the analytical prediction (Eq.~(\ref{eq:Hp})). }
(c)~The jumping height as a function of the rescaled substrate diameter, $d/r_c$, for the same loading conditions ($(R, h) = (25, 2.0)$~mm, $e\simeq38$~mm, $mg = 0.45$~N).
The blue shaded region is the critical diameter for the contact geometry transition $d = 2 \alpha_r r_c$ {(Eq.~(\ref{eq:ecrc}))}. The dashed line represents the analytical formula for the jumping height on the narrow substrate (Eq.~(\ref{eq:ecrc}), $d/2r^*\ll1$). {We plot the corresponding experimental data (empty square) for the flat substrate $2r^*/d\to0$.}
}
\label{Fig5}
\end{figure}

\subsection{Maximum jumping height on the narrow substrate}\label{sec:Hn}

We measure the maximum jumping height of the popper, $H$, for various sets of shell geometry, $(h,R)$, Young's modulus, $E$, net weight, $mg$, and the initial displacement of the apex $e$. The experimental parameters are coordinated by our analytical prediction, Eq.~\eqref{eq:Hp}, in Fig.~\ref{Fig5}(a). We find that the Eq.~\eqref{eq:Hp} (dashed line) is in excellent agreement with experimental data without any fitting parameters. The possible origin of their difference is the relation between the compressive pressure $p^*$ and the apex displacement, $e$ (Eq.~\eqref{eq:pst}), where the dynamics of the air is neglected. Although more sophisticated (computational) modeling by taking into account the fluid-structure interactions will improve the predictions, we leave the extension as a future study. The excellent agreement between the experimental results and theory indicates that the elasticity and geometry of the point-indented pressurized shell are crucial in predicting the dynamics of the popper. 

\subsection{Jumping phase diagram}\label{sec:JumpNot}

We have shown that Eq.~\eqref{eq:Hp} provides sufficient prediction for the maximum jumping height of the popper. We further apply Eq.~\eqref{eq:Hp} for another experiment, where we uncover whether the popper can jump or not. According to Eq.~\eqref{eq:Hp}, $H_{\rm point}$ increases monotonically with $e$, i.e., the popper can jump higher as we indent the apex largely. In other words, when the apex displacement is small, the stored elastic energy would be insufficient to launch the popper. The condition that the popper can jump or not corresponds to that $H_{\rm point}$ is larger than the initial displacement, $e$, or not; $H_{\rm point} = e$. The critical condition for the pros and cons of the jump is 
\begin{eqnarray}
    \frac{\beta}{3\{12(1-\nu^2)\}^{3/4}}\frac{Eh^{5/2}e^{1/2}}{R} = mg.\label{eq:Jcond}
\end{eqnarray}
We have performed the systematic experiment by varying the net weight $mg$ to identify the critical condition. We classify the experimental outcomes where the popper can/cannot leave the substrate as \textit{Jump}/\textit{No Jump}, respectively, in Fig,~\ref{Fig5}(b). We find that Eq.~\eqref{eq:Jcond} is in excellent agreement with the experimental data, indicating again that the shell elasticity is essential to predict the jumping performance of the popper. 

\subsection{Critical diameter for narrow/flat substrate}\label{sec:dia}

The jumping phenomena of the popper on the sufficiently narrow substrate are now entirely predictable by applying the elasticity and geometry of the shell. Here, we discuss the effect of cylinder diameter, $d$, on the maximum jumping height, $H$. We perform the jumping experiment to measure $H$ for various values of $d$, while we set $(h, R) = (25, 2.0)$~mm, $mg = 0.45$~N, and $e\simeq38$mm (Fig.~\ref{Fig5}(c)). We find that the maximum height of the popper increases as we decrease the diameter of the substrate ($d\to 0$), i.e., the popper jumps higher on the narrow substrate than on the flat one, which is consistent with our daily observations. Interestingly, the maximum height significantly increases once the diameter falls behind the critical length scale, which is close to the critical diameter for the contact geometry transition, $2r^*$ (Eq.~\eqref{eq:ecrc_s}).

We can now argue the role of the rigid cylinder in the jumping phenomena of the elastic shell. When the diameter of the cylinder is sufficiently large, the apex of the shell cannot come into contact with the substrate prior to the contact geometry transition. When the contact radius, which is equal to the radius of the buckled rim, $r$, reaches the critical radius $r^*$, the shell snaps and pops out. However, when the diameter of the substrate is sufficiently smaller than the critical radius, $d\ll 2r^*$, the cylinder is always in contact with the apex (the buckled rim is free), and the radius of the rim, $r$, decreases to zero without passing the contact geometry transition. Hence, the narrow substrate prohibits the contact geometry transition. 
When the diameter is of the order of the critical radius, $d\sim2r^*$, one may need to take into account the geometry of the top surface of the cylinder. Indeed, it is reported that the contact geometry is crucial in the snap-buckling of the shell via axisymmetric numerical simulation~\cite{Huang2024}. Although we speculate that the maximum height, $H$, follows the logistic curve-like function of $2r*/d$, we leave the prediction for the intermediate regime $d\sim2r^*$ as an intriguing future theoretical study, which is beyond the scope of this paper.


\section{Discussion and Summary}\label{sec:dis}

In this paper, we have studied the jumping dynamics of elastomeric shells on a rigid substrate. We experimentally analyze the jumping performance on the rigid cylinder of diameter, $d$. In contrast to the jumping dynamics of the popper on the flat substrate~\cite{Abe2025}, the popper does not generate impulsive force upon jumping. Despite the absence of the impulsive force, the popper can offer sufficient jumping power. We have derived the analytical formula for the maximum jumping height on the sufficiently narrow substrate ($d\to0$). The maximum jumping height results from the complex interplay of elasticity, geometry, and contact mechanics, highlighted by the nontrivial relations between experimental parameters. Our prediction is in excellent agreement with the experimental data, not only for the maximum height but for the jumping ability of the popper. Lastly, we systematically vary the diameter of the cylinder to clarify the critical condition of the substrate being sufficiently narrow or flat. We identify that when the diameter is close to the critical diameter for the contact geometry transition of the shell in contact with the flat substrate.

Our model experimental results successfully validate the proposed analytical framework (Eqs.~(\ref{eq:Hflat}) and (\ref{eq:Hp})). We now examine whether our results are still useful in the behavior of the popping toy (\textit{e.g.,} Fig.~\ref{Fig1}). Given that our popping toys are sufficiently light and pressure difference is not present, it would be allowed to take the limit of $mg\to0$ and keep the leading order terms of Eqs.~(\ref{eq:Hflat}) and (\ref{eq:Hp}) only. 
The maximum height on the flat substrate obeys the scaling as $H_{\rm flat}\simeq H_{\rm s}\sim Eh^4/mgR$, while that on the narrow substrate scales as $H_{\rm point}\sim Eh^{5/2}e^{3/2}/mgR$. When the popper is fully inverted before the jump, the apex displacement is approximately written as $e\approx 2R$. The ratio of the maximum jumping height between flat and point-like substrate obeys the following scaling relation:
\begin{eqnarray}
    \frac{H_{\rm point}}{H_{\rm flat}} \sim \left(\frac{R}{h}\right)^{3/2},\label{eq:HpHf}
\end{eqnarray}
implying that the thinner shells would jump higher on the narrow substrate than on the flat one. In our demonstration, we have used the popper toys of $(R,h) \approx (20, 2)$~mm. After computing all the dimensionless prefactors ($\alpha_F,\alpha_e$ are set to be the median in the prediction range), we get the ratio between ${H_{\rm point}}$ and ${H_{\rm flat}}$ as ${H_{\rm point}}/{H_{\rm flat}}\approx1.4$, roughly consistent with our demonstration. Although we need to be careful about the validity of the scaling relation, Eq.~(\ref{eq:HpHf})\footnote{For example, the popping toy does not necessarily jump vertically due to the contact geometry in the initial configuration. The reproducibility in experiments for the popping toy is not sufficient for further validation.}, we believe that our predictive framework for the jumping behavior could be extended to more complex popping robots.

We note the difference between fundamental mechanisms behind the jumping ability on the flat and narrow substrates from the bifurcation point of view. The contact geometry transition plays a crucial role in jumping on the flat substrate, which is characterized by the hysteretic response in the indentation test (Fig.~\ref{Fig4}(a)). The force-displacement curve exhibits a subcritical nature, where the loading and unloading processes track different paths. The shell is indented to store the elastic energy, which is released upon jumping. The indenting and jumping processes correspond to the loading and unloading protocols, respectively. The difference in the loading/unloading curves, i.e., the area enclosed in the cyclic test, represents the energy dissipation in the jumping process on the flat substrate (highlighted by the snapping sound). However, as discussed in the main text, when the shell snaps on the narrow substrate, the contact geometry transition does not exist. In contrast to the flat-indentation test, the force-displacement curve by the point-indentation exhibits almost no hysteresis (Fig.~\ref{Fig4}(b)), where there is no subcritical bifurcation. Given that the loading and unloading process follows nearly identical paths, the energy dissipation is less pronounced upon jumping. In other words, the jumping process on the narrow substrate utilizes the stored elastic energy efficiently. Hence, the jumping height on the narrow substrate observed in our experiment is higher than that on the flat one.

We have revealed that jumping on the point-like substrate is more efficient than that on the flat substrate, at least with respect to the maximum height. Although jumping on the narrow substrate is energetically efficient, the impulsive force triggered by snap-buckling does not exist (See Fig.~\ref{Fig3}). The jumping process on the narrow-like substrate is similar to the spring-like actuation~\cite{Patek2023}. The switching mechanism between snap-buckling and spring-like actuation is mediated by contact geometry between the shell and substrate. We expect that the role of contact geometry uncovered in this study would be applied in optimizing the actuation mechanism of soft and flexible robots in the near future. For example, by optimizing the surface geometry of the shell, one may be able to control the jumping performance by design~\cite{Qiao2020}.


\begin{acknowledgments}
This work was supported by MEXT KAKENHI 24H00299 (T.G.S.), JST FOREST Program, Grant Number JPMJFR212W (T.G.S.). 

\end{acknowledgments}

\bibliography{apssamp}

\end{document}